\documentclass[sigconf]{acmart}

\usepackage{booktabs}
\usepackage{multirow}
\usepackage{array}
\usepackage{amsmath}
\usepackage{subfigure}
\usepackage{tablefootnote}
\usepackage{makecell}

\newcolumntype{x}[1]{>{\let\newline\\\arraybackslash\hspace{0pt}}p{#1}}
\newcolumntype{y}[1]{>{\centering\let\newline\\\arraybackslash\hspace{0pt}}p{#1}}

\newcommand {\mymarginpar}[1]{\marginpar{#1}}
\renewcommand {\marginpar}[1]{}

\newcommand {\rfig}[1]{\autoref{fig:#1}}

\newcommand {\bsec}[2]{\section{#1}
                       \label{sec:#2} }

\newcommand {\rsec}[1]{\autoref{sec:#1}}

\newcommand {\bsubsec}[2]{\mymarginpar{sec:#2}
                       \subsection{#1}
                       \label{sec:#2} }

\newcommand {\rsubsec}[1]{\autoref{sec:#1}}

\newcommand {\beq}[1]{
                      \begin{equation}
                      \label{eq:#1} }
\newcommand {\eeq}{\end{equation}}
\newcommand {\beqno}[1]{\begin{eqnarray}
                      \nonumber}

\newcommand {\eeqno}{ && \end{eqnarray}}
\newcommand {\req}[1]{Eq.~(\ref{eq:#1})}

\newcommand {\bear}[1]{
                       \begin{eqnarray}
                       \label{eq:#1} }

\newcommand {\bearno}[1]{
                       \begin{eqnarray}
                       \nonumber}

\newcommand {\eear}{\end{eqnarray}}
\newcommand {\eearno}{\end{eqnarray}}

\newcommand {\btab}[1]{
                       \begin{table}
                       \centering
                       \begin{tabular}{#1}}
\newcommand {\etab}[3] {
                       \end{tabular}
                       \caption[#3]{#2}
                       \label{tab:#1}
                       \end{table}
                       \vspace{.1in}}
\newcommand {\rtab}[1]{\autoref{tab:#1}}

\newcommand {\btabular}[1]{\begin{center}
                       \begin{tabular}{#1}}
\newcommand {\etabular}{\end{tabular}
                       \end{center}}

\newcommand {\bdefin}[1]{\begin{definition}\label{def:#1}}
\newcommand {\edefin}       {\end{definition}}

\newcommand {\bpro}[1]{\begin{property}
                      \label{pro:#1} }
\newcommand {\epro}   {\end{property}}

\newcommand {\bprop}[1]{\begin{proposition}
                      \label{prop:#1} }
\newcommand {\eprop}       {\end{proposition}}

\newcommand {\blem}[1]{\begin{lemma}
                      \label{lem:#1}}
\newcommand {\elem}   {\end{lemma}}

\newcommand {\bthe}[1]{\begin{theorem}
                      \label{the:#1} }
\newcommand {\ethe}   {\end{theorem}}

\newcommand {\bcor}[1]{\begin{corollary}
                      \label{cor:#1} }
\newcommand {\ecor}   {\end{corollary}}

\newcommand{\hide}[1]{}

\AtBeginDocument{%
  \providecommand\BibTeX{{%
    \normalfont B\kern-0.5em{\scshape i\kern-0.25em b}\kern-0.8em\TeX}}}
    
\copyrightyear{2023}
\acmYear{2023}
\setcopyright{acmlicensed}\acmConference[KDD '23]{Proceedings of the 29th ACM SIGKDD Conference on Knowledge Discovery and Data Mining}{August 6--10, 2023}{Long Beach, CA, USA}
\acmBooktitle{Proceedings of the 29th ACM SIGKDD Conference on Knowledge Discovery and Data Mining (KDD '23), August 6--10, 2023, Long Beach, CA, USA}
\acmPrice{15.00}
\acmDOI{10.1145/3580305.3599845}
\acmISBN{979-8-4007-0103-0/23/08}
\begin{document}

\title{Impact-Oriented Contextual Scholar Profiling\\using Self-Citation Graphs}


\author{Yuankai Luo}
\affiliation{%
  \institution{Beihang University}
   \streetaddress{37 Xueyuan Road}
 \city{Beijing}
 \country{China}
  \postcode{100191}
}
\email{luoyk@buaa.edu.cn}

\author{Lei Shi}
\authornote{Corresponding authors.}
\affiliation{%
  \institution{Beihang University}
   \streetaddress{37 Xueyuan Road}
 \city{Beijing}
 \country{China}
  \postcode{100191}
}
\email{leishi@buaa.edu.cn}

\author{Mufan Xu}
\affiliation{%
  \institution{Beihang University}
   \streetaddress{37 Xueyuan Road}
 \city{Beijing}
 \country{China}
  \postcode{100191}
}
\email{16231231@buaa.edu.cn}

\author{Yuwen Ji}
\affiliation{%
  \institution{Beihang University}
   \streetaddress{37 Xueyuan Road}
 \city{Beijing}
 \country{China}
  \postcode{100191}
}
\email{18241019@buaa.edu.cn}

\author{Fengli Xiao}
\affiliation{%
  \institution{Beihang University}
   \streetaddress{37 Xueyuan Road}
 \city{Beijing}
 \country{China}
  \postcode{100191}
}
\email{zy2206114@buaa.edu.cn}

\author{Chunming Hu}
\affiliation{%
  \institution{Beihang University}
   \streetaddress{37 Xueyuan Road}
 \city{Beijing}
 \country{China}
  \postcode{100191}
}
\email{hucm@buaa.edu.cn}

\author{Zhiguang Shan}
\authornotemark[1]
\affiliation{%
  \institution{State Information Center}
   \streetaddress{58 Sanlihe Street}
 \city{Beijing}
 \country{China}
  \postcode{100045}
}
\email{shanzg@sic.gov.cn}

\renewcommand{\shortauthors}{Yuankai Luo et al.}

\begin{abstract}
Quantitatively profiling a scholar's scientific impact is important to modern research society. Current practices with bibliometric indicators (e.g., h-index), lists, and networks perform well at scholar ranking, but do not provide structured context for scholar-centric, analytical tasks such as profile reasoning and understanding. This work presents GeneticFlow (GF), a suite of novel graph-based scholar profiles that fulfill three essential requirements: structured-context, scholar-centric, and evolution-rich. We propose a framework to compute GF over large-scale academic data sources with millions of scholars. The framework encompasses a new unsupervised advisor-advisee detection algorithm, a well-engineered citation type classifier using interpretable features, and a fine-tuned graph neural network (GNN) model. Evaluations are conducted on the real-world task of scientific award inference. Experiment outcomes show that the F1 score of best GF profile significantly outperforms alternative methods of impact indicators and bibliometric networks in all the 6 computer science fields considered. Moreover, the core GF profiles, with 63.6\%$\sim$66.5\% nodes and 12.5\%$\sim$29.9\% edges of the full profile, still significantly outrun existing methods in 5 out of 6 fields studied. Visualization of GF profiling result also reveals human explainable patterns for high-impact scholars.
\end{abstract}


\maketitle

\vspace{-0.15 in}
\bsec{Introduction}{Intro}

\begin{figure*}[t]
\centering
\vspace{-0.13 in}
\includegraphics[width=0.85\linewidth]{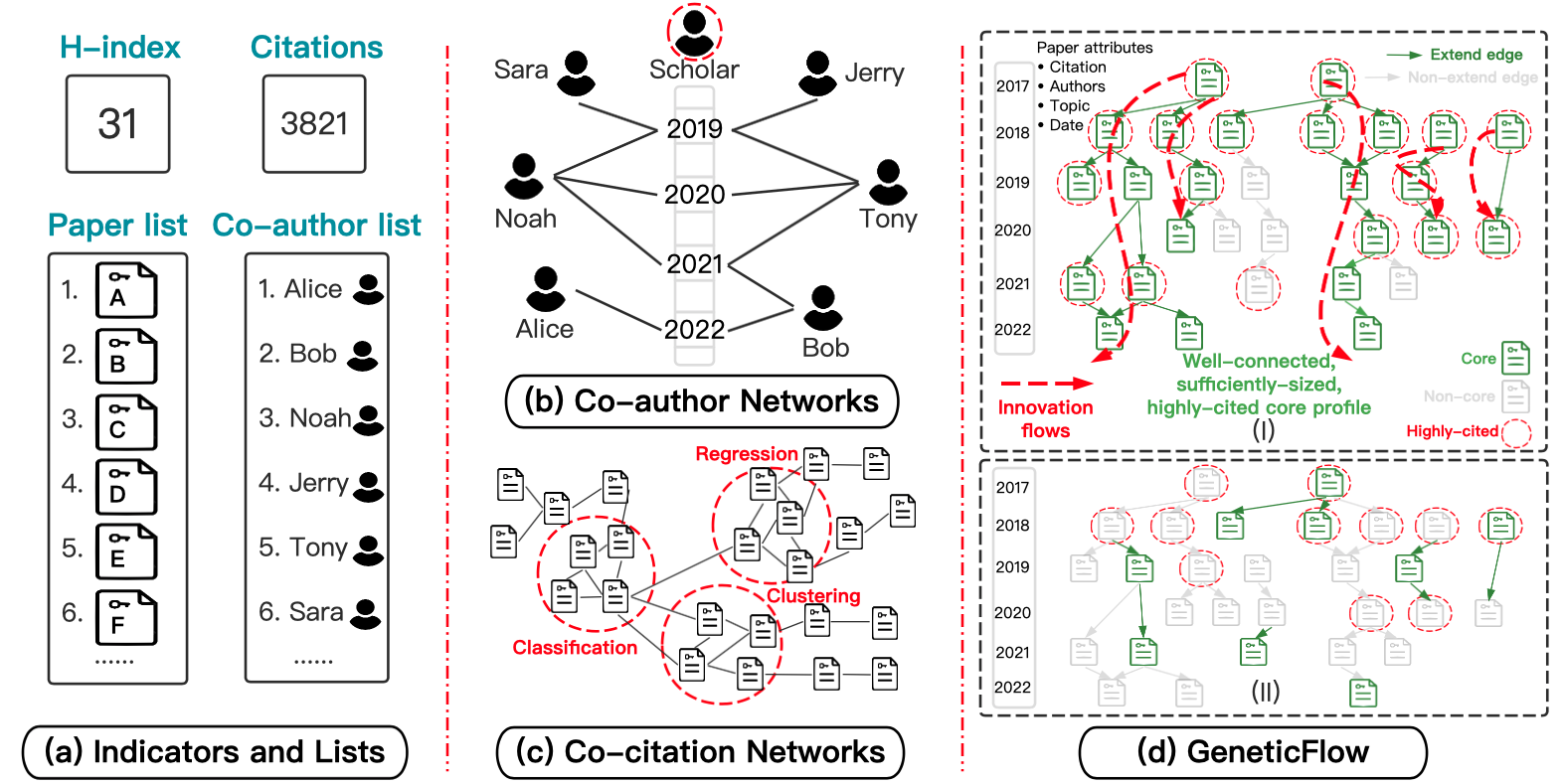}
\vspace{-0.15 in}
\caption{GeneticFlow profiling design in (d).I for a scholar with higher impact than the scholar in (d).II,  versus classical scholar impact indicators in (a) and bibliometric networks in (b)(c).}
\vspace{-0.20 in}
\label{fig:Overview}
\end{figure*}




Data-driven scholar profiling, promoted by the practice of IT giants via Google Scholar \cite{GoogleScholar} and Microsoft Academic Search \cite{MAS}, has drawn significant attention of the research community \cite{DBLP}\cite{ResearchGate}\cite{gasparyan2017researcher}\cite{yuan2018survey}. Commonly, scholar profiling is composed of two steps: 1) accurate extraction of 360-degree academic demographics of a scholar from the web, which is normally considered an information retrieval task; and 2) appropriate aggregation, analysis, and representation of the extracted academic data, which is largely a data mining task. This work focuses on the latter step and formulates the \textbf{impact-oriented scholar profiling problem} -- how to arrange a scholar's academic data to best represent his/her scientific impact. Here the impact is defined as the breadth and depth of one's scientific contribution, as well as their community recognition. The problem is related to important applications of scholar profiles, such as serving possible references for academic award selection and tenure evaluation \cite{schimanski2018evaluation}, and predicting one's future academic output \cite{fortunato2018science}.


Existing scholar profiling websites generally design two types of views. First, the standard \emph{indicator and list} view exemplified by Google Scholar, ACM author profile \cite{ACMAuthorProfile} and AMiner profile \cite{AMiner}, as shown in \rfig{Overview}(a), provides key bibliographic indicators (\# of citations, h-index \cite{hirsch2005index}, etc.) as an overview of the scholar. The view is complemented with multiple lists as the context (publications, co-authors, etc.). The indicators are believed to be helpful for scholar ranking, though the supplemented paper/co-author lists are normally unprocessed. There is a missed opportunity to exploit the structure within these lists for impact-oriented contextual tasks beyond ranking, such as reasoning and understanding of scholar profiles. Second, academics websites often exhibit various \emph{bibliometric networks}. These networks mitigate the drawback of indicator and list by illustrating the relational context of a scholar. Yet, most existing bibliometric networks are not designed for impact-oriented scholar profiling. For example, the co-authorship network delineates one's external connections over time (\rfig{Overview}(b)), but is short in representing a scholar's internal scientific contribution mostly designated by his/her core publications. The co-citation \cite{small1973co} network is useful for detecting emerging topic clusters (\rfig{Overview}(c)), but is not suitable for scholar-centric profiling, especially at revealing the evolution of one's research impact.

Our work goes beyond the basic task of scholar ranking and considers impact-oriented contextual tasks of scholar profile reasoning, understanding, and analysis. Three requirements should be met on the new profiling problem: \textbf{structured-context (R1)}, the complex academic data of a single scholar, reflecting one's temporal, topical, and relational context, should be integrated into a structured representation to ensure the efficiency and effectiveness of scholar profiling; \textbf{scholar-centric (R2)}, the profile should focus on the target scholar only, excluding the entities attributed to others, e.g., his/her co-authors. It should also be impact-oriented, omitting ordinary features such as affiliation, address, and email; \textbf{evolution-rich (R3)}, the profile should track the evolution of a scholar's scientific impact, presumably with richer context than a standard timeline chart of citation counts.


%

The main idea of this work is to capitalize on the structured nature of graph-based representation, while satisfying the above requirements on scholar profiling. We introduce GeneticFlow (GF), a suite of scholar-centric graph representations serving as their contextual profile. To obtain structured context (R1), GF exploits self-citation relationship, which is previously deemed as detrimental but now shown quite effective in profiling the innovation flows of a scholar. To achieve scholar-centric (R2) and evolution-rich (R3) qualities, GF supports data-driven profiling on both graph nodes and edges to extract representative components of a scholar's scientific impact. Finally, GF is capable to differentiate among scholars that are not set apart by standard approaches. For example, \rfig{Overview}(d).I and II illustrate the GF profile of two fictional scholars having the same citations, h-index, and paper count. The scholar on the top is analytically of higher impact than the one on the bottom, with a well-connected, sufficiently-sized, and highly-cited core GF profile in the foreground (see \rsubsec{Case} for more details).

Computing, analyzing, and evaluating GF over real-world big scholar data poses nontrivial challenges, for which we have made the following contributions:

\vspace{-0.05 in}
\begin{itemize}
 \item We design the GF concept and introduce a complete framework to construct them for any scholar. To profile GF nodes, a scholar's core papers are inferred to be those most representative to his/her scientific impact. For GF edges, the extend-type citations are extracted as graph cores to represent the evolution of a scholar's scientific contribution. The design is empirically shown as effective by visualizations of GF profiling result. Case studies on the NLP field also demonstrate key GF patterns on high-impact scholars.

 \item We propose an unsupervised advisor-advisee detection method based on new, interpretable definition on co-authorship ties. The method helps to efficiently detect core papers of a scholar over big data, without the use of any training label. Experiment results over OpenReview data validate its effectiveness in comparison to baseline methods.

 \item We introduce an optimized classifier to detect extend-type citations of a scholar. To resolve label sparsity and data unbalance issue, a professional data annotation process as well as relevant tool support are developed and conducted. By jointly applying bibliometric, temporal, and content features crafted for extend-type citations, our classifier greatly outperforms all previous methods in detection results.

 \item We conduct quantitative evaluation by inferring major scientific award recipients (e.g., ACM fellows) with GNN representation of GF profiles. A new benchmark dataset is built
 from big scholar data sources with millions of academic records. On all the 6 key fields studied, the best GF profile significantly outperforms classical indicator-based and network-based methods on award inference performance (F1). Further experiment reveals the importance of detected core GF profiles for the inference.
 \end{itemize}
\vspace{-0.05 in}
Data and code is available at: \url{https://github.com/visdata/GeneticFlow/}

\vspace{-0.05 in}
\bsec{Preliminaries}{Graph}
\vspace{-0.03 in}
\bsubsec{GeneticFlow Framework}{GeneticFlow}


As shown in \rfig{Overview}(d), the full GF profile of a scholar in the background is a timed, directed acyclic graph composed of all the papers authored by this scholar with impact-oriented paper attributes (citations, topic, etc.), and self-citations among these papers. This graph focuses on the scholar-centric impact.
Notably, GF exploits the well-known self-citation relationship that is conventionally deemed as useless or even detrimental, because authors have the incentive to boost self-citations \cite{fowler2007does}. Yet, we affirm that, besides the role of increasing impact, self-citations also serve an important function of delineating the innovation flows of a scholar. Using self-citations, the plain list of scholar publications is orchestrated into structured graph representation for rich-context profiling (R1).

The basic GF design still leaves several gaps toward impact-oriented scholar profiling: a) how to detect the set of core papers to construct a compact scholar profile? 
While generic paper ranking metrics pile up in the literature \cite{kanellos2019impact}, the core papers defined in our context are those most representative to the scientific impact of the target scholar (wrt. R2); b) how to detect the set of self-citations that truly represent the evolution of a scholar's research contribution (wrt. R3)? As complained in the literature, excessive and artificial self-citations could be issued for nonscientific reasons \cite{szomszor2020much}.

\vspace{-0.12 in}
\bsubsec{Impact-Oriented Scholar Profiling Problem}{Problem}


Formally, under the framework of GeneticFlow, the \emph{full GF profile} of a scholar $s$ is defined by a graph $G(s)=\{V(s),E(s)\}$ where $V(s)$ denotes the set of $n$ paper nodes authored by $s$ and $E(s)$ denotes the set of $m$ reversed self-citation edges among $V(s)$. Each node $v_i$ is associated with a timestamp $t$ (pub. year by default), an ordered list of authors $A=\{a_1,a_2,...\}$, and an extra set of paper attributes $\Phi$. Each edge $e_{ij}=(v_i,v_j)$ denotes the citation from paper $v_j$ to $v_i$, i.e., the reversed citation influence link.

As the full GF profile can be deformed by nonessential publications and citations, the impact-oriented scholar profiling problem is defined as finding the subgraph $G^*(s)$ of $G(s)$ that best represents the impact of scholar $s$, as illustrated by the foreground graphs of \rfig{Overview}(d). They are called \emph{core GF profiles}. The profiling problem is further decomposed into two sub-problems: a) \textbf{node profiling} that detects the set of core papers $V^*(s) \subseteq V(s)$ published by the scholar $s$, and the core paper attributes $\Phi^* \subseteq \Phi$ that contributes to the profiling (the paper timestamp is already included); b) \textbf{edge profiling} that detects the set of core citation edges $E^*(s) \subseteq E(s)$ that represents the evolution of the scholar's scientific contribution.




\vspace{-0.05 in}

\bsec{Related Work}{Related}
\bsubsec{Author-Level Impact Indicators}{Indicator}


In a comprehensive review, Wildgaard et al. examined 108 author-level impact indicators and classified them into 5 categories \cite{wildgaard2014review}. The first type involves indicators of individual's publication counts, mostly the raw number of papers and patents. A variant of this type considers the different credits from a publication received by all the authors, according to author order and class \cite{waltman2015field}\cite{moya2013research}. The publication count based indicators do not explicitly measure scientific impact of each paper because they do not value the quality of an author's research output. The second type of indicators are designed to evaluate authors by the venues (journals) they published at, thus are essentially journal-level indicators \cite{haustein2012multidimensional}. The third indicator type focuses on the quality of research output, predominately by paper citations. Typical indicators are the sum of all paper citations, the average number of citations per paper, and the number of highly cited papers. For example, Google Scholar introduces i10-index, the number of papers with at least 10 citations. Google displays it as a major index in the scholar profile \cite{GoogleScholar}. The limitations of the third-type indicators lie in that they do not explicitly count the paper quantity and do not consider the dynamic evolution of paper citation counts, which could be important for a scholar’s impact.

The fourth type is ranking-based author indicators that only count top-impact publications, notably the widely adopted h-index metric \cite{hirsch2005index}. In its original form, Hirsch defined the h-index to be the maximum value of $h$ such that the author has published at least $h$ papers with at least $h$ citations each. The main advantage of h-index is that the indicator strikes a balance between the quantity and quality of an author's publication. From its invention, a huge amount of follow-up researches were conducted to propose variants or extensions of h-index. For example, one class of research argued that once a paper is selected for h-index computation (called Hirsch core \cite{rousseau2006new}), its number of citations is not used anymore. Solutions include g-index \cite{egghe2006theory}, A-index \cite{jin2006}\cite{rousseau2006new}, and $h_{\alpha}$-index \cite{van2008generalizing}, which give more weights to highly cited papers in the Hirsch core. The major deficiency of the h-index class for our problem lies in that, though composite with both quantity and quality of author impact, these indices are single indicators which can not reveal the full context of author-level impact. This is extremely important for contextual scholar profiling tasks beyond ranking. The last author-level indicator type involves metrics that reflect the dynamics of author impacts over time. A typical indicator is AR index that takes into account both h-index like scores and publication age \cite{jin2007}.

\vspace{-0.16 in}

\bsubsec{Bibliometric Networks}{Network}


Bibliometric networks describe the relationship among one or more types of academic entities (author, paper, venue, etc.) \cite{yan2012scholarly}\cite{eck2014visualizing}\cite{kong2019academic}. These networks are generally built from the bibliographic information available at academic data sources such as Microsoft Academic Graph (MAG) \cite{MAG}, DBLP \cite{DBLP}, CiteSeerX \cite{CiteSeerX}, SemanticScholar \cite{SemanticScholar}, etc. Mainstream bibliometric networks include co-authorship networks, citation networks, and keyword co-occurrence networks. On citation networks focused in this work, two well-known definitions of co-citation \cite{small1973co} and bibliographic coupling \cite{kessler1963bibliographic}, are often applied to construct networks featuring topic-based paper clusters, also known as the research front \cite{chen2006citespace}. Our study is different from these topic-level profiling approaches in that we focus on the representation of author-level impact.


The historiographs generated by HistCite software \cite{garfield2002algorithmic}\cite{garfield2003we}\cite{garfield2004historiographic} are also related to the proposed GF profiling method. By considering the publication date of each paper,  historiograph visualization becomes a hierarchical flow graph over time. Yet, historiograph is designed to capture the evolution of a scientific topic/field and the papers included in a graph are not limited to those of a single scholar. Instead, GF is only composed of the papers published by this scholar, and moreover, our approach focuses on the papers and citations representative for the scholar's scientific impact. In a similar study, Hellsten et al. \cite{hellsten2007self} highlighted the importance of self-citations to reveal the topic evolution of a scholar. However, their self-citation networks were customized to detect a scientist's mobility in research fields. Again, the importance of papers and citations to the author-level impact is not considered. Another thread of relevant work proposed algorithms to summarize the full citation network into compact graph abstractions for analysis \cite{shi2015vegas} and visualization \cite{huang2019eiffel}. These approaches generally work on paper-centric citation networks, but not the scholar-centric network. Also, the summarization loses context and detail that are important for impact-oriented profiling.

\vspace{-0.05 in}

\bsec{Contextual Scholar Profiling}{Model}

\begin{table}[t]
\centering
\setlength{\tabcolsep}{2pt}
\caption{Academic data sources used in this work.}
\vspace{-0.13 in}
\label{tab:DataSource}
\begin{tabular}{@{}cccccccc@{}}
\toprule
\makecell{Data\\source} & \makecell{\# of\\paper} & \makecell{\# of\\author}  & \makecell{\# of\\citation}  & \makecell{Author\\order}  & \makecell{Citation\\context}  & \makecell{Topic\\info}  & \makecell{Version\\date}            \\ \midrule
MAG & 237M & 240M & 1.6B & Yes & Few & Yes & 202005 \\
ARC & 73K & 59K & 1M & Yes & Yes & N/A & 202107 \\
OpenAlex & 243M & 213M & 1.7B & Partial & Few & Yes & 202210 \\ \bottomrule
\end{tabular}
\vspace{-0.15 in}
\end{table}

\bsubsec{Data Source and Pre-processing}{Data}


\textbf{Big scholar data from MAG.} The proposed GF profiling method is mainly applied to MAG data \cite{MAGDownload}, which covers 237M papers from all science areas, 240M authors, and 1.63B citations (\rtab{DataSource}). MAG is now the largest open academic data source and is more comprehensive than alternatives \cite{visser2021large}, e.g., DBLP (no citation links), AMiner (fewer citations). Importantly, on the Computer Science (CS) area focused in this work, MAG provides most required data to build GF profile: paper title, abstract, authors, pub. date, topics, and citation links. There is only one deficiency: merely 9.9\% citations have their context available, i.e., sentences containing each citation, which could contain important features for citation classification. We rectify this by merging MAG with detailed yet smaller-scope data sources such as ACL Anthology Reference Corpus (ARC) \cite{ARC}.




\textbf{Citation context augmentation.} The ARC data has a good coverage on papers from the natural language processing (NLP) field. ARC includes both the metadata (title, authors, date, abstract, etc.) of NLP papers and their full-text download. We parse these full-text data with ParsCit \cite{ParsCit} to extract all the citation context and additional citation links missing in MAG. By matching ARC and MAG on paper title and year, 84.1\% ARC papers till 2019 can be correctly linked to an MAG entry. On major NLP venues (ACL, EMNLP, NAACL, COLING, TACL), the matching rate reaches 98.3\%.



\vspace{-0.10 in}
\bsubsec{Node Profiling}{NodeRe}


We adopt two basic rationales in profiling the core papers representative to a scholar: a) the paper should be of high impact; b) the scholar should make significant contribution to the paper. While the first rationale is largely met by including citation info. in the paper node attribute of GF, most algorithmic effort in node profiling focuses on the latter rationale to determine the author contribution. In fact, the traditions of representing author contribution are versatile, and in some venues, explicit acknowledgement of individual author's contribution is even required. Our method is based on two assumptions generally followed in the CS area.

\noindent \textbf{Assumption 1 (author order)}: A paper's contribution is unequally credited to all authors by author order unless the paper is alphabetically ordered.

\noindent \textbf{Assumption 2 (advisor-advisee credit sharing)}: An author's contribution to the paper is also credited to his/her advisor if only: a) the advisor is a co-author of the paper; and b) the advisor-advisee relationship is active at the publication date of the paper.

\noindent \textbf{Theorem (author contribution)}: On any paper $v$ published at time $t$, the probability for the $k$th author $a_k$ to contribute significantly can be estimated by
\beq{Contribution}
p_{cont}(a_k) = max(\frac{1}{k},~~ \underset{\forall l \neq k}{max}\frac{p_{AA}(a_k, a_l, t)}{l})
\eeq
Here the popular harmonic credit allocation scheme \cite{hagen2009credit} is adopted that the $k$th author takes a credit of $1/k$. $p_{AA}(a_k, a_l, t)$ denotes the probability of $a_k$ being the advisor of $a_l$ at time $t$.

The GF node profiling method then decomposes into two steps. First, all the scholars with a high rate of alphabetically ordered papers are detected and avoided in applying GF profiling. Second, the papers significantly contributed by the target scholar are extracted by the above theorem. The key is to detect time-sensitive advisor-advisee relationship involving the target scholar.



\textbf{Alphabetical authorship detection.} Over MAG/ARC datasets, we manage to detect sub-fields of CS with significantly more alphabetically ordered papers. The scholars focused in these fields are avoided in GF profiling because it is hard to identify core papers using their nominal author order. The challenge here lies in that alphabetical authorship is not often tagged in the paper, nor are these acknowledgements easy to extract and parse. We propose a statistical method based on hypothesis testing to estimate the chance of a field using alphabetical authorship. The null hypothesis is set for a field as not using any alphabetical authorship. The percentage of papers with their authors listed exactly the same with alphabetical order ($p_{a-order}$) is used as the test statistic. Under the null hypothesis, it is reasonable to assume all authors to be ordered by their contribution, independent with their names. Then the theoretical average of test statistic can be computed by
\beq{alphabetical}
p_{a-order} = \sum_{\substack{k=1}}^\infty \frac{N_k}{N} \cdot \frac{1}{k!}
\eeq
where $N$ indicates the number of all papers in a field, $N_k$ indicates the number of papers with exactly $k$ authors, $k!$ gives the number of possible ways to place all the $k$ authors of a paper. There is only one alphabetical author order for a paper with $k$ authors, while there are $k!$ possible author orders. This probability $1/k!$ is further weighted by the ratio of papers with $k$ authors, which gives \req{alphabetical}.

We apply the hypothesis test to 8 CS sub-fields in our dataset. In each field, the papers authored by top-500 scholars are considered. 
By examining the bias of observed test statistics from those by null hypothesis, we find that the fields of TCS and PL are quite different from the other fields, with large increases in the observed $p_{a-order}$. These two fields are then excluded from GF profiling because of the large number of alphabetically ordered papers there. The details results can be found in Appendix~\ref{ap-aa}.


\begin{table*}[t]
\centering
\vspace{-0.10 in}
\setlength{\tabcolsep}{3pt}
\caption{Performance of extend-type citation inference using various classifiers, feature sets, and the comparison with literature.}
\vspace{-0.13 in}
\label{tab:ExtendPerf}
\begin{tabular}{@{}cccccccccc@{}}
\toprule
\multirow{2}{*}{Metric} & \multicolumn{3}{c}{Classifier} & \multicolumn{4}{c}{Ablation study}                     & \multicolumn{2}{c}{Previous result}                                                                                                                                                                         \\ \cline{2-10}
                        & Extra-trees   & MLP    & DNN   & (-) Paper-meta & (-) Cite-net & (-) Temporal & (-) Content & \cite{teufel2006automatic}\cite{valenzuela2015identifying}\cite{jurgens2018measuring} merged & Report in \cite{jurgens2018measuring} \\ \midrule
F1 score                & \textbf{.646$\pm$.014}          & .543$\pm$.018   & .544$\pm$.015  & .636$\pm$.007          & .639$\pm$.010        & .639$\pm$.005        & \textbf{.471$\pm$.009}       & \textbf{.418$\pm$.019}                                                                                                                                            & .403$\pm$.029                                                   \\
AUC                     & \textbf{.902$\pm$.005}          & .806$\pm$.016   & .785$\pm$.014  & .871$\pm$.009          & .898$\pm$.006        & .899$\pm$.005        & \textbf{.796$\pm$.008}       & \textbf{.841$\pm$.009}                                                                                                                                               & .775$\pm$.017 \\
ACC                & \textbf{.924$\pm$.002}          & .901$\pm$.004   & .899$\pm$.004  & .921$\pm$.002          & .922$\pm$.001        & .924$\pm$.002        & \textbf{.895$\pm$.002}       & .949$\pm$.001                                                                                                                                               & \textbf{.976$\pm$.001} \\
\bottomrule
\end{tabular}
\vspace{-0.13 in}
\end{table*}

\textbf{Advisor-advisee detection.} The problem has been previously studied in the literature \cite{liu2019shifu2}\cite{zhao2018identifying}\cite{wang2010mining}. However, most algorithms are supervised in that they depend on labeled real-life advisor-advisee dataset. According to GF profiling goal, we are concerned with not only formal advisor-advisee ties in the university, but also informal relationship such as mentor-mentees within/across institutions between senior and junior researchers, for which labeled data is scarce. Moreover, the only existing unsupervised method, TPFG by Wang et al. \cite{wang2010mining}, is based on the Kulczynski measure not designed and interpretable for advisor-advisee detection.

In this work, we propose a new advisor-advisee detection method starting from an intrinsic definition of generic advisor-advisee relationship. The method is unsupervised, fully interpretable, and fast enough to be applied on MAG.

\noindent \textbf{Definition (advisor-advisee relationship)}: The advisor of an advisee in a research field at time $t$ is characterized as \emph{an experienced researcher in the field} (D1),  who supervised \emph{a sufficient number and ratio of major papers by the advisee} (D2) in \emph{a sufficiently long time} (D3) on the \emph{early career of the advisee in the field} (D4).

The algorithm following the definition can be summarized as
\beq{advisor}
p_{adr}(a_k, a_l, t) = \frac{N_{a_k}(0,t) - N_{a_k, a_l}(0,t)}{N_{a_k, a_l}(0,t)}
\eeq
\beq{advisee}
p_{ade}(a_k, a_l, t) = \max_{\substack{t_{0} \leq t \leq t_{1},~~t_{1} -  t_{0} \geq S_{len}\\numerator \geq S_{adr}}}\frac{\underset{t_{0} \leq t \leq t_{1}}{\sum}\hat{N}_{a_k, a_l}(t) Mod(t)}{\hat{N}_{a_l}(t_{0}, t_{1})}
\eeq
\beq{AA}
p_{AA}(a_k, a_l, t) = min(1.0, p_{adr}(a_k, a_l, t)) \times min(1.0, p_{ade}(a_k, a_l, t))
\eeq

In \req{advisor}, the competency for $a_k$ being an advisor of  $a_l$ at time $t$ is inferred by the ratio of the number of papers published solely by the advisor ($N_{a_k}(0,t)-N_{a_k, a_l}(0,t)$) to the number of papers on this advising relation ($N_{a_k, a_l}(0,t)$), both before time $t$. A larger ratio indicates a more experienced advisor (D1). The advising papers in the denominator are those with advisee ranked as top-3 authors and advisor ranked after the advisee.

In \req{advisee}, the probability for $a_l$ being an advisee of $a_k$ at time $t$ is inferred by the maximal percentage of advisee's major papers supervised by that advisor (D2), during a sufficiently long time period $[t_0, t_1]$ (D3). The number of major papers ($\hat{N}_{a_l}$) is counted by only considering those with the advisee as the first, second, or third author, using the weights of 1, $\frac{1}{2}$, $\frac{1}{3}$ by the harmonic allocation scheme. To satisfy D2 and D3, we also require the numerator of \req{advisee} and the length of period to be lower bounded by pre-set parameters, namely $S_{adr}$ and $S_{len}$ respectively. During the counting of advising papers ($\hat{N}_{a_k, a_l}$), each paper count is further multiplied by a career modifier ($Mod(t)$) to satisfy D4. When the advisee publishes in the field for a sufficiently long time or has produced a good number of papers, $Mod(t)$ starts to decay exponentially from an initial value of 1. This will prevent the false positive ties detected between two senior scholars collaborating on every paper during some period. Finally, both advisor and advisee probabilities are upper bounded by a maximal probability of 1, and then combined as the advisor-advisee rate (\req{AA}).


\begin{table*}[t]
\centering
\small
\caption{The list of 20 significant, independent features (Ft) used for the inference of extend-type citations.}
\vspace{-0.13 in}
\label{tab:EdgeFeature}
\begin{tabular}{@{}cccx{0.5\textwidth}cc@{}}
\toprule
Category                             & Name                     & \#Ft & \multicolumn{1}{c}{Description} & Sig. & Dataset                     \\ \midrule
\multirow{3}{*}{Paper-meta}      & \# of citations\_cited & 1          & citation count of the cited paper &  0.0016      & MAG                         \\
                                     & year\_diff               & 1          & publication year difference between cited and citing papers &  0.00027     & ARC \& MAG                  \\
                                     & \# of shared\_authors               & 1          & the number of shared authors between cited and citing papers          &  1.9e-48       & ARC \& MAG                  \\ \midrule
\multirow{2}{*}{Cite-net}    & co-citation              & 2          & co-citation metrics between cited and citing papers          &  $\leq$1.0e-07      & \multirow{2}{*}{ARC \& MAG} \\
                                     & bib-coupling             & 1          & bibliographic coupling metrics between cited and citing papers          &  5.2e-08     &                             \\ \midrule
Temporal          & cross-correlation        & 3          & cross-correlations between citation time series of cited/citing papers         &  $\leq$0.037     & MAG                         \\ \midrule
\multirow{5}{*}{Content} & content-similarity       & 1          & cosine similarity between vectorized content of cited and citing papers          &  1.3e-16 & \multirow{5}{*}{ARC}        \\
                                     & \# of cite\_occurrences       & 1          & the number of total occurrences of in-text citations of this citation link          &   4.0e-09      &                             \\
                                     & \# of cites\_occur\_sec       & 3          & \# of cite\_occurrences in key sections          &   $\leq$0.044      &                             \\
                                     & cite\_relative\_pos      & 4          & position of in-text citations in paper, section, sub-sec., sentence &   $\leq$0.049    &                             \\
                                     & lexical\_pattern         & 2          & appearance of certain phrases: ``an/the extension'', ``our previous'', etc.         &  $\leq$5.0e-11        &                             \\ \bottomrule
\end{tabular}
\vspace{-0.15 in}
\end{table*}

\bsubsec{Edge Profiling}{EdgeRe}




Our method to detect core citations leverages the existing studies on citation type classification \cite{teufel2006automatic}\cite{valenzuela2015identifying}\cite{jurgens2018measuring}\cite{tuarob2019automatic}. While taxonomies on citation type vary a lot, they are consistent on the most important type, namely \emph{extend-type citations}.
It is typically defined as, e.g. in \cite{teufel2006automatic}, ``the author uses cited work as basis or starting point''. When a scholar extends the previous paper of himself/herself, the new work will probably be an evolution of the scholar's research ideas. The evolution, together with the citation impact of each paper within the evolution, can be important to his/her scientific impact. Our GF profile faithfully accomplishes the “evolution-rich” requirement by modeling the profile as a temporal graph (best revealed in \rfig{Overview}(d).I). We believe, satisfying the “evolution-rich” requirement plays a key role in the success of GF profiling method.

To infer extend-type citations, supervised learning techniques can be applied. Initially, we managed to merge three existing annotated datasets on NLP to obtain labels for supervision  (Teufel et al. 2006 \cite{teufel2006automatic}, Valenzuela et al. 2015 \cite{valenzuela2015identifying}, Jurgens et al. 2018 \cite{jurgens2018measuring}). As these previous works were positioned for all-type citation classifications with extend type as the rarest class, only 89 extend-type citation labels are present, with respect to the other 1604 non-extend citation labels. For effective feature extraction, we have excluded the citation type labels on which either citing or cited paper is not in the NLP-ARC dataset. The initial dataset achieves a rather low F1 score (0.418) on the detection of the extend class (\rtab{ExtendPerf}), because of the highly unbalanced nature and small positive label size.

We then augment the dataset by manually annotating extend-type citations. 
In our practice, both interactive tool support and a standard annotation process are established to ensure the accuracy. The details of the annotation work can be found in Appendix~\ref{ap-a}. Finally, 133 new extend-type labels are obtained, leading to a dataset of 222/1604 positive/negative extend-type citation samples.

We mitigate interpretability concerns on citation classification by hand-crafting four categories of raw features interpretable for extend-type citation inference. As shown in \rtab{EdgeFeature}, the first category includes metadata of cited and citing papers, e.g., \# of citations, sharing of authors (self-cite), and the difference in publication year. The second category is features extracted from their citation networks, notably the co-citation and bibliographic coupling metrics. The third category is temporal correlation measures between the citation count time series of citing and cited papers. The last category contains content and lexical patterns extracted from the citation context and full text. 
These features are evaluated based on their significance in differentiating extend-type and non-extend citations in the labeled dataset. 
Finally, 20 features are selected (\rtab{EdgeFeature}), all having $p$ values smaller than 0.05 under the Mann-Whitney U test.

On the selection of classifiers, most latest machine learning techniques have been examined. The results with representative classifiers are summarized in \rtab{ExtendPerf}. In a 10-fold cross-validation setting, the Extra-Tree model \cite{geurts2006extremely} is shown to be the best (F1: 0.646, AUC: 0.902), significantly better than the Multilayer Perceptron (MLP) with a single hidden layer (F1: 0.543, AUC: 0.806) and the DNN model with 10 layers (F1: 0.544, AUC: 0.785). The performance is also much better than applying the same classifier to the initial merged dataset with 89 extend labels (F1: 0.418, AUC: 0.841), and the model reported in Jurgens et al. \cite{jurgens2018measuring} (F1: 0.403, AUC: 0.775). Note that the overall accuracy (ACC) is better in previous models. It is because the previous models are applied to highly unbalanced data with fewer positive labels, with more non-extend citations easier to classify. Ablation studies for feature importance were also conducted. As shown in \rtab{ExtendPerf}, the content features extracted from the citation context are the most useful among all feature types.

On the NLP-ARC dataset, our calibrated classifier is applied to all the self-citations of top-500 scholars. 18.1\% citation links are classified as extend-type. On MAG datasets, there is neither extend-type labels nor content features. The model on NLP-ARC re-trained without content feature is applied to these datasets as is. 


\vspace{-0.10 in}
\bsubsec{Representation Learning}{GraphRe}

\begin{table*}[t]
\centering
\vspace{-0.05 in}
\setlength{\tabcolsep}{2pt}
\caption{Experiment data from 8 CS sub-fields. 50 true award recipients and 150 other scholars are sampled in each field.}
\vspace{-0.13 in}
\label{tab:FieldEva}
\begin{tabular}{@{}cy{0.32\textwidth}cccccc@{}}
\toprule
CS & Awards & \# of awardees & Sample  & \multicolumn{2}{c}{Full GF profile: \# of nodes, edges} & \multicolumn{2}{c}{Core profile} \\ \cline{5-8}
                         sub-field     &   (except ACM fellow \& Turing award)      &    (top-500 scholars)                  &  list                        & Awarded (50)          & Others (150)           & ~~Nodes~~            & ~~Edges~~            \\ \midrule
NLP-ARC   &   ACL Lifetime Achievement Award / Fellow &   77 &   \#1$\sim$\#207   &  121$\pm$56,205$\pm$173 &   93$\pm$50,153$\pm$134  &     66.5\%  &  12.8\% \\
Database        &   SIGMOD Innovations Award    &   114 &   \#1$\sim$\#247  &   118$\pm$61,166$\pm$126          &   74$\pm$36,112$\pm$79    &   64.0\%          & 12.8\% \\
Security        &   SIGSAC Outstanding Innovation Award &   81  &   \#1$\sim$\#208  &   138$\pm$79,190$\pm$167  &   123$\pm$66,145$\pm$105   &  65.5\%  & 18.3\% \\
DM              &   SIGKDD/ICDM Innovations/Research Award  &   108 &   \#1$\sim$\#235  & 169$\pm$136,305$\pm$392     &   133$\pm$66,233$\pm$181      &  65.9\% & 25.1\% \\
HCI             &   SIGCHI Lifetime Research Award / Academy    &   117     &   \#1$\sim$\#251  &   113$\pm$61,160$\pm$145  &   99$\pm$51,135$\pm$94          &  63.8\%  &  29.9\%  \\
SE                           & SIGSOFT Outstanding Research Award                    & 56                                   & \#1$\sim$\#369                   & 81$\pm$41,86$\pm$85         & 69$\pm$35,67$\pm$52   & 63.6\%  & 12.5\% \\
TCS                           & SIGACT Donald E. Knuth Prize                    & 127                                   & \#1$\sim$\#239                   & 114$\pm$47,215$\pm$170          & 99$\pm$43,202$\pm$150          &  N/A           & N/A           \\
PL          & SIGPLAN PL Achievement Award                    & 135                                   & \#1$\sim$\#244                   & 90$\pm$34,165$\pm$134          & 87$\pm$37,187$\pm$180          & N/A            & N/A
\\ \bottomrule
\end{tabular}
\vspace{-0.1 in}
\end{table*}


To apply GF methods to downstream tasks, we introduce graph neural network (GNN) models to learn high performance representation of profiling results. The general design of our model follows classical GNN architecture. The core component is a tandem of graph convolution layers (filters), where each layer learns and updates the representation of every node by propagating and aggregating their neighborhood information. In the literature, different kinds of graph convolution filters have been proposed \cite{ChebNet}\cite{GCN}\cite{GAT}\cite{ARMA}. All these filters can be applied in the GF framework and have also been evaluated in our experiment (\rsec{Eva}). Optionally, after each graph convolution, a pooling operation \cite{zhang2019hierarchical}\cite{EdgePool} can be conducted to make graphs smaller for better node representation. After the convolution layers, the learned node representations are collapsed via a readout layer to construct the final representation of the entire graph. A few fully connected MLP layers are added in the end, as proxy for downstream classification or regression tasks.

\textbf{Node attribute embedding.} To represent a scholar's impact comprehensively, the GNN model on GF profiles is designed to embed 4 node attributes: the paper's total citation count, the publication date, the scholar's order in the paper, and the paper's topic vector. In MAG, the topic information is provided for each paper as a list of related topic categories, namely the field of study. All MAG topic categories form a hierarchical graph with up to 6 levels, where CS is one of 19 level-1 topic categories. We extract a sub-graph from the MAG topic hierarchy with the CS category as root. The MDS algorithm \cite{MDS} is applied to this sub-graph to compute a low-dimensional embedding for each topic category belonging to the CS field. On each paper, the embedding vectors of all its related topic categories are averaged to generate the final topic embedding of the paper. On the topic vector embedding, we also consider OpenAlex \cite{OpenAlex}, the successor to MAG after its recent retirement. 
In \rsubsec{SA-result}, ablation study results reveal the effective attribute profiling in GF with citation and pub. date. The MAG topic vector is shown to be better than its OpenAlex replacement.

\vspace{-0.05 in}
\bsec{Evaluation}{Eva}
\bsubsec{Scientific Award Inference}{SA-result}

\textbf{Data and task.} To validate the effectiveness of GF profiling, we consider the task of inferring major scientific award recipients. A new benchmark dataset is established, as shown in \rtab{FieldEva}, which includes 8 sub-fields of CS: NLP, Database (DB), Security, Data Mining (DM), Human-Computer Interaction (HCI), Software Engineering (SE), Theoretical Computer Science (TCS), and Programming Language (PL). In each field, we only consider the highest-class technical achievement/innovation awards, plus the title of ACM fellow and Turing award as general awards in CS. ACM fellows are qualified as awardees in a field only if s/he has published a sufficient number of papers in the field. The 8 sub-fields are selected due to the higher number of award recipients than the other CS sub-fields ($>50$ within top-500 scholars), thus alleviates the data imbalanced issue for training. The TCS and PL fields are later excluded due to their significant use of alphabetical authorship (\rsubsec{NodeRe}). 
The dataset of each field is built from MAG by default, except that the NLP dataset (NLP-ARC) is synthesized from MAG and ARC, with 30\% more citation links than the original MAG data and additional citation context features.

In each CS sub-field, we sample 200 scholars including 50 award recipients and 150 other scholars. The inference task is a typical supervised learning that predicts awardees out of all the 200 labeled scholars in each field. The inference-by-field design is adopted as the award-impact relationship is highly likely to be much different across research fields. To ensure that the sampled awardees and unawardees do not vary significantly in basic scholarly profiles, we rank the scholars by their h-index in the field and only select top scholars for inference. To achieve this, a minimal sampling list is computed for each field, e.g., \#1$\sim$\#207 in NLP-ARC. The list guarantees that at least 50 award recipients and 150 other scholars are included. The final samples are then obtained through two uniform samplings on awardee and unawardee groups in the list separately. The statistics of GF profiles in both groups are given in \rtab{FieldEva}. The profiles in the awardee group have a higher number of nodes/edges, but the difference is not large enough for inference due to high variance within each group. Note that the sample size of 200 in each field seems to be small compared with the large number of total scholars. Yet, the current samples have already covered almost 50\% award recipients in all the 6 CS sub-fields considered. Further raising the sample size can only be done by mostly adding negative samples, boosting the already high data imbalance ratio. This may not be useful for the evaluation of profiling method.

\begin{figure}[t]
\centering
\includegraphics[width=0.9\linewidth]{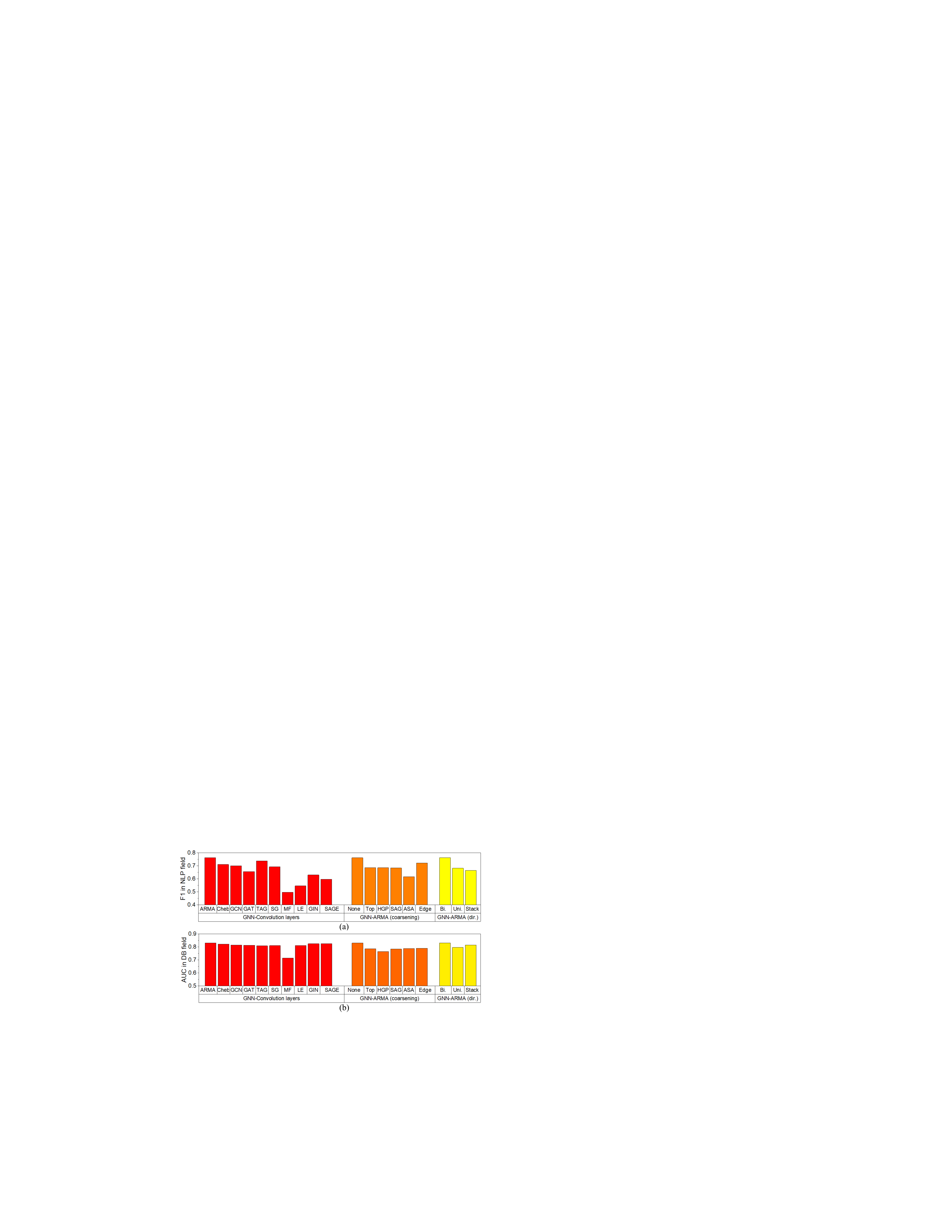}
\vspace{-0.18 in}
\caption{GNN performance on the award inference task (F1,AUC@awarded class): (a) NLP field; (b) DB field.}
\vspace{-0.20 in}
\label{fig:GNNPerf}
\end{figure}


\begin{table*}[t]
\centering
\vspace{-0.10 in}
\setlength{\tabcolsep}{2pt}
\caption{F1 measure in the award inference task using GeneticFlow and alternative methods.}
\vspace{-0.13 in}
\label{tab:FieldResult}
\begin{tabular}{@{}cccccccccc@{}}
\toprule
CS        & \multicolumn{2}{c}{GeneticFlow}           & \multicolumn{4}{c}{Author-Level Impact Indicators}                    & \multicolumn{3}{c}{Bibliometric Networks}                            \\ \cline{2-10}
sub-field & Full profile            & Best core profile         & SVM             & XGB             & RF              & MLP             & CC              & BC              & CA                         \\ \midrule
NLP-ARC       & \textbf{.762$\pm$.016 (p<1e-4)} & .720$\pm$.018 (p=2e-4) & .632$\pm$.012 & \textbf{.636$\pm$.013} & .621$\pm$.019 & .629$\pm$.016 & .531$\pm$.030 & \textbf{.578$\pm$.021} & .473$\pm$.034  \\
Database  & .634$\pm$.018 (p=0.034) & \textbf{.638$\pm$.016 (p=0.012)}  & .517$\pm$.020 & \textbf{.546$\pm$.021} & .526$\pm$.020 & .517$\pm$.016 & .550$\pm$.021           & \textbf{.588$\pm$.012}           & .501$\pm$.035                  \\
Security  & \textbf{.606$\pm$.020 (p=0.044\tablefootnote{With 20\% edges removed.})} & .551$\pm$.022 & .557$\pm$.025 & .572$\pm$.016 & .548$\pm$.018 & \textbf{.589$\pm$.021} & \textbf{.576$\pm$.017}           & .572$\pm$.018  & .528$\pm$.021             \\
DM        & \textbf{.653$\pm$.020 (p=0.007)} & .627$\pm$.014 (p= 0.045) & \textbf{.590$\pm$.012} & .533$\pm$.018 & .574$\pm$.018 & .574$\pm$.016 & .563$\pm$.022           & \textbf{.569$\pm$.019}           & .476$\pm$.020               \\
HCI       & \textbf{.644$\pm$.018 (p=1e-4)} & .625$\pm$.016 (p=0.001) & \textbf{.562$\pm$.011} & .558$\pm$.017 & .548$\pm$.025 & .528$\pm$.017 & \textbf{.551$\pm$.024}           & .527$\pm$.022           & .466$\pm$.029           \\
SE        & .665$\pm$.011 (p=0.023) & \textbf{.668$\pm$.009 (p=0.014)} & \textbf{.596$\pm$.011} & .558$\pm$.016 & .512$\pm$.020 & .593$\pm$.014 & \textbf{.607$\pm$.023}           & .595$\pm$.019           & .523$\pm$.028         \\
\bottomrule
\end{tabular}
\vspace{-0.1 in}
\end{table*}

\begin{table}[t]
\centering
\small
\setlength{\tabcolsep}{2pt}
	\caption{Ablation study result by the GNN-ARMA model.}
\vspace{-0.15 in}
\label{tab:AblationStudy}
\begin{tabular}{@{}ccccccc@{}}
\toprule
 \makecell{Field\\(F1)}   & \makecell{(-)\\Graph}          & \makecell{(-)\\\# of Cites}     & \makecell{(-)\\Date}    & \makecell{(-)\\Order}       & \makecell{(-)\\Topic}          & \makecell{(+)\\OpenAlex} \\ \midrule
NLP (.762) & .671$\pm$.026 & .655$\pm$.007 & .455$\pm$.017 & .758$\pm$.022 & .755$\pm$.017 & .755$\pm$.018 \\
DB (.634) & .455$\pm$.024 & .618$\pm$.011 & .586$\pm$.024 & .611$\pm$.021 & .609$\pm$.014 & .614$\pm$.014 \\
\bottomrule
\end{tabular}
\vspace{-0.15 in}
\end{table}

\textbf{Model setup.} On GF method, both full profile and core profile are evaluated as predictive features, using GNN representation models. The core profile contains the paper nodes significantly contributed by the target scholar ($p_{cont}>0.5$, see \rsubsec{NodeRe}) and the self-citation edges with top extend-type probability (\rsubsec{EdgeRe}). Ten latest GNN convolution layers and 5 optional graph coarsening methods are implemented with the standard PyG library \cite{PyG}. For comparison, three popular author-level indicators (\# of papers, \# of citations, h-index) and three author-level bibliometric networks are also introduced as predictors for inference. The indicators are concatenated and evaluated with four high performance classifiers (SVM, XGBoost (XGB), Random Forest (RF), MLP). The networks include co-citation (CC) and bibliographic coupling (BC) having the same set of paper nodes with full GF profile, and the co-authorship network (CA) having all co-authors of the target scholar as network nodes. To make fair comparison, the node attributes on GF profile (e.g., paper citation count) are also included in these three bibliometric networks. GNN models are also applied to the inference with bibliometric networks.

The experiment with each model and field setting is conducted with a stratified 10-fold cross-validation to accommodate the imbalanced data. The cross-validation is repeated 10 times to test model stability. On performance metrics, we are mainly concerned with the optimal F1 of the awarded (minority) class. Occasionally, AUC is also reported to assess the overall minority class performance.


\textbf{Result.} \rfig{GNNPerf} compares the result of different GNN models in NLP and DB datasets. Among convolution layer (red bars of \rfig{GNNPerf}), the latest ARMA model \cite{ARMA} (leftmost) performs the best on both F1 and AUC and is chosen as the default convolution layer for GF. The trade-offs using graph coarsening (orange bars) and different edge directions (yellow bars) are also evaluated. Graph coarsenings do not improve the model for our task, but bring large complexity overhead. Hence, no graph coarsening is applied. Treating profile as bi-directional graphs yields the best performance.


Ablation study is conducted to investigate the importance of each type of GF features for the task. It is shown in \rtab{AblationStudy} that removing graph structure\footnote{We use the features on all paper nodes w/o edges to proceed the inference task.}, paper citation count, and pub. date will lead to considerable performance degradation ($>0.1$ in F1) in either NLP or DB field. Yet, removing paper topic, author order, or replacing with OpenAlex topic hierarchy only introduce mild effect ($\leq0.025$ in F1) on both fields tested. The low data quality of MAG/OpenAlex's topic tag could be an explanation. This result recommends the best attribute profiling setting for GF.



\begin{figure*}[t]
\centering
\vspace{-0.10 in}
\includegraphics[width=0.83\linewidth]{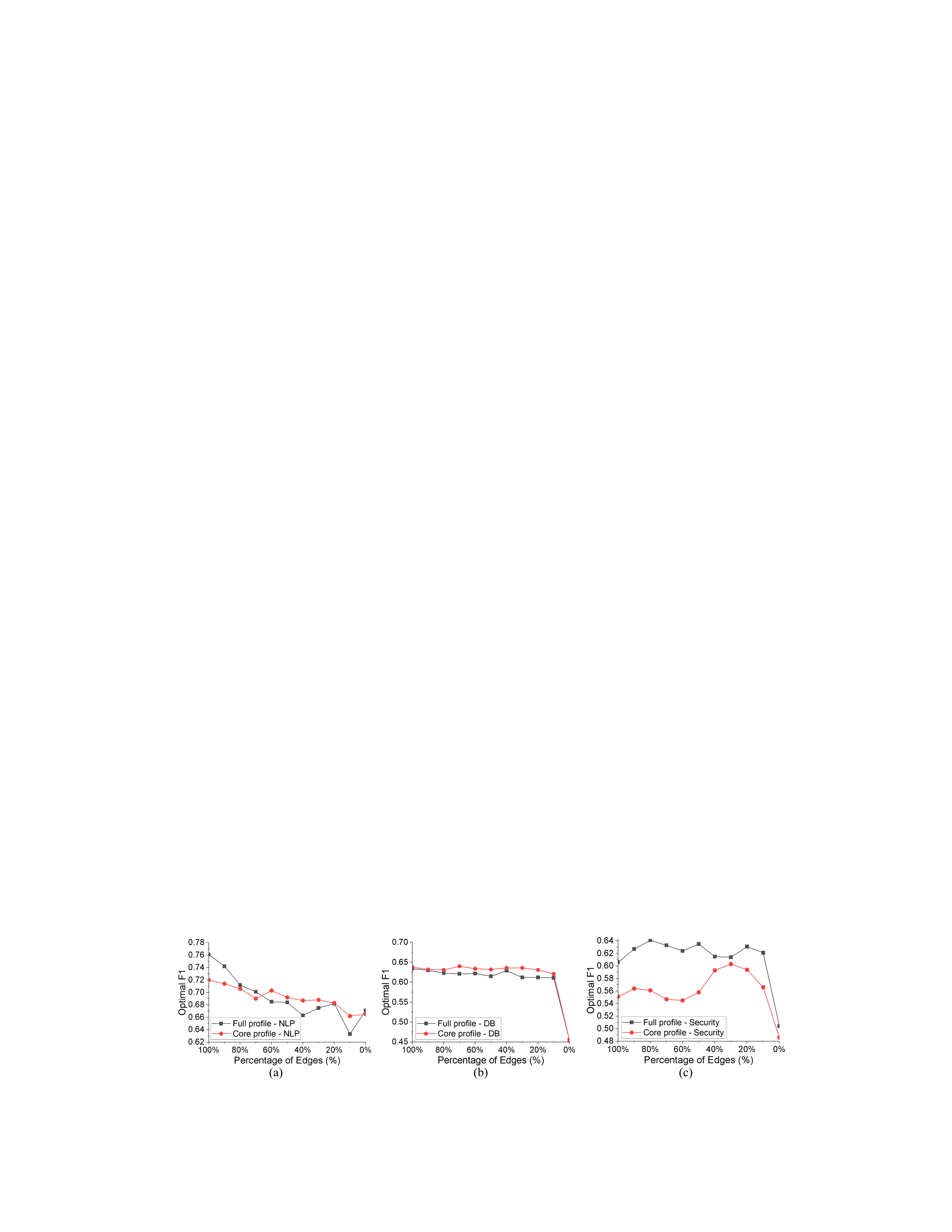}\\
\includegraphics[width=0.83\linewidth]{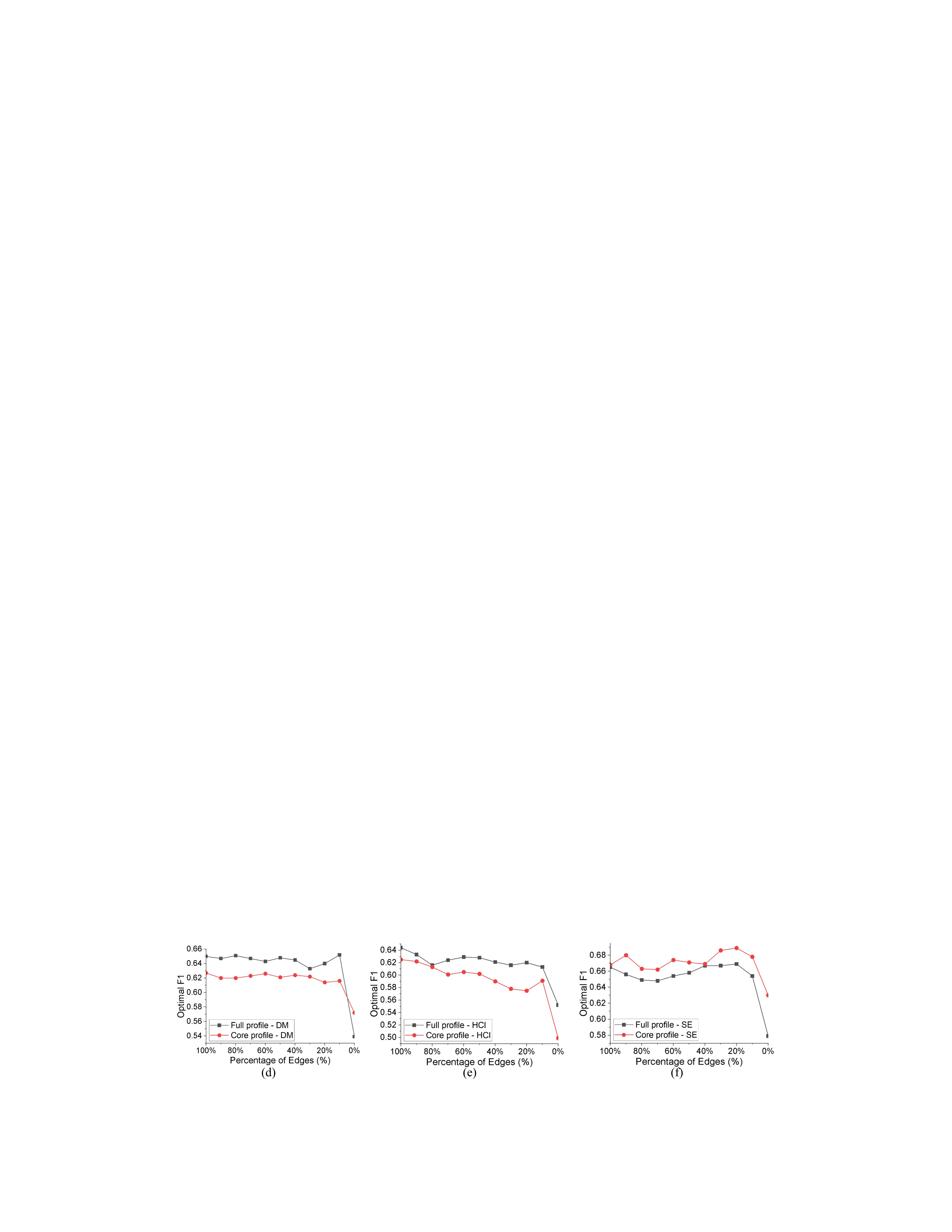}
\vspace{-0.17 in}
\caption{The performance of full/core GF profiles with varying edge percentages in 6 CS sub-fields.}
\vspace{-0.1 in}
\label{fig:EdgePerf}
\end{figure*}

The GF method is then compared with existing author-level indicators and networks. Using the full GF profile, as shown in \rtab{FieldResult}, the inference performance by F1 is much better than the best indicators and networks, for all 6 fields studied. The advantage is significant\footnote{We use a confidence interval of 95\% unless otherwise noted.} in 5 fields ($p\leq0.023$) except the Security field. When 20\% edges are removed from the full profiles of Security field, a significant advantage is also achieved over the baseline indicators/networks ($p=0.044$,  see \rfig{EdgePerf}(c)).





We follow up to study the effectiveness of node profiling in GF methods. In \rtab{FieldResult}, it is shown that the F1 measure by the best core profile is worse than its full profile in 4 fields, and better in the other 2 fields. However, the differences between core and full profile are not significant in all fields studied. Moreover, compared with the best indicator/network, the best core GF profile still achieves significant advantage ($p\leq0.045$) in all fields except Security. In the Security field, the F1 of core profile with 20\%$\sim$40\% edges is better than all the baselines.

On the effect of edge profiling, we study the performance of core GF profiles with varying percentages of top self-citation edges used. As shown by the red+circle lines of \rfig{EdgePerf}, in all fields, F1 finally drops as more edges are removed, though some local increases are also observed. This demonstrates the usefulness of self-citations in GF profiling for award inference. Notably, the performance degradation becomes much quicker when the top 10\%$\sim$20\% edges start to be removed, indicating the higher importance of the most probable extend-type citations. Using 50\% self-citation edges, sufficiently good performance can be achieved in 5 fields except Security, with significantly higher F1 than the best author-level indicator. We also evaluate the full GF profile with edges incrementally removed. As shown by the black+rectangle lines of \rfig{EdgePerf}, the F1 curve is quite similar to those of core profiles.

\textbf{Discussion.} The experiment results imply several findings. First, on award inference task, the proposed GF method greatly outperforms existing alternatives, e.g., author-level indicators/networks. F1 measure of the best GF profile is significantly higher than that of the best alternative in all 6 CS sub-fields studied. Even with the default full GF profile, significant advantage can be achieved in 5 out of 6 fields.



Second, the extracted core papers and extend-type citations are shown to be key components of GF profile. On node profiling, though removing non-core nodes often decreases the performance, the difference is not significant in all fields, with the benefit of using more compact profile for user analysis. Moreover, the best core profile is significantly better than all indicators/networks in 5 out of 6 fields. On edge profiling, the top 10\%$\sim$20\% self-citation edges are shown to be the most important to GF profiling. Once removed, inference performance drops sharply. These edges correspond to extend-type citations, which reach a percentage of 18.1\% in the NLP-ARC dataset (\rsubsec{EdgeRe}). In fact, the GF method provides quite focused scholar profiling result. As listed in the last column of \rtab{FieldEva}, we compute the smallest core GF profile in each field that is significantly better than existing methods\footnote{For the Security field, an advantage is achieved in average F1 without significance.}. These compact profiles only require 63.6\%$\sim$66.5\% nodes and 12.5\%$\sim$29.9\% edges of the full scholar-centric graph.


Finally, on GF profiling for award inference, we observe certain difference across research fields and datasets applied. On NLP field studied with ARC+MAG data, because of the more complete citation links and accurate extend-citation model, the inference performance is much higher than the other fields using MAG data. This is shown by the smallest p-value in \rtab{FieldResult} ($\leq2e-4$ for full/core profiles). Meanwhile, the Security field seems to be an outlier in applying GF, where significant advantage can hardly be observed. We analyzed the award data in Security field. Because of the relatively young history of the field, within all the 81 awardees there, only 12 ever received ``Innovation Award'' in Security and 69 (85.2\%) are ACM Fellows without a field award. A large portion of these ACM Fellows could be named due to the contribution in other fields and moved to Security after that. In comparison, only 9.1\%(14.5\%) awardees in NLP(HCI) fields are ACM Fellows without field award. A future work would be validating this hypothesis by removing migrating scholars from the awarded group.
\vspace{-0.1 in}
\bsubsec{Advisor-Advisee Detection}{AA-result}

\begin{table}[t]
\centering
\small
\setlength{\tabcolsep}{3pt}
\caption{Advisor-advisee detection on OpenReview data.}
\vspace{-0.15 in}
\label{tab:AA-data}
\begin{tabular}{@{}ccccccc@{}}
\toprule
\multirow{2}{*}{Field} & \# of Positive & \# of Negative &  \multicolumn{2}{c}{Ours}         & \multicolumn{2}{c}{TPFG} \\ \cline{4-7}
&  labels  & labels &F1 &ACC & F1   & ACC \\
\midrule
NLP & 365 & 116 & 0.880 & 0.830 & 0.856 & 0.771\\
DM & 145 & 35 & 0.900 & 0.844 & 0.879 & 0.794\\
\bottomrule
\end{tabular}
\vspace{-0.15 in}
\end{table}

To evaluate advisor-advisee detection algorithms, we build a real-life dataset from the OpenReview website \cite{OpenReview}, on which scholar users sometimes provide their advisor information. The NLP and DM fields popular on OpenReview are considered, where the personal websites of top-200 scholars in each field and their co-authors are investigated. In total, 365/116 and 145/35 positive/negative advisor-advisee pairs, as shown in  \rtab{AA-data}, are obtained in the NLP and DM fields respectively. The data crawling detail and its usage for parameter calibration of our advisor-advisee detection algorithm can be found in Appendix~\ref{ap-b}.

On the OpenReview data, both our new advisor-advisee detection algorithm and the baseline TPFG model are applied. Note that due to complexity issue, the academic social network used in TPFG is limited to the top scholars in our dataset and all their co-authors. As demonstrated in \rtab{AA-data}, in both NLP and DM fields, our proposed algorithm achieves high scores on F1 (0.88, 0.9) and ACC (0.83, 0.844). In comparison, the baseline TPFG obtains worse performance on both F1 (-0.021$\sim$-0.024) and ACC (-0.05$\sim$-0.059).

\vspace{-0.10 in}
\bsubsec{GF Visualization}{Case}

\begin{figure}[t]
\centering
\includegraphics[width=\linewidth]{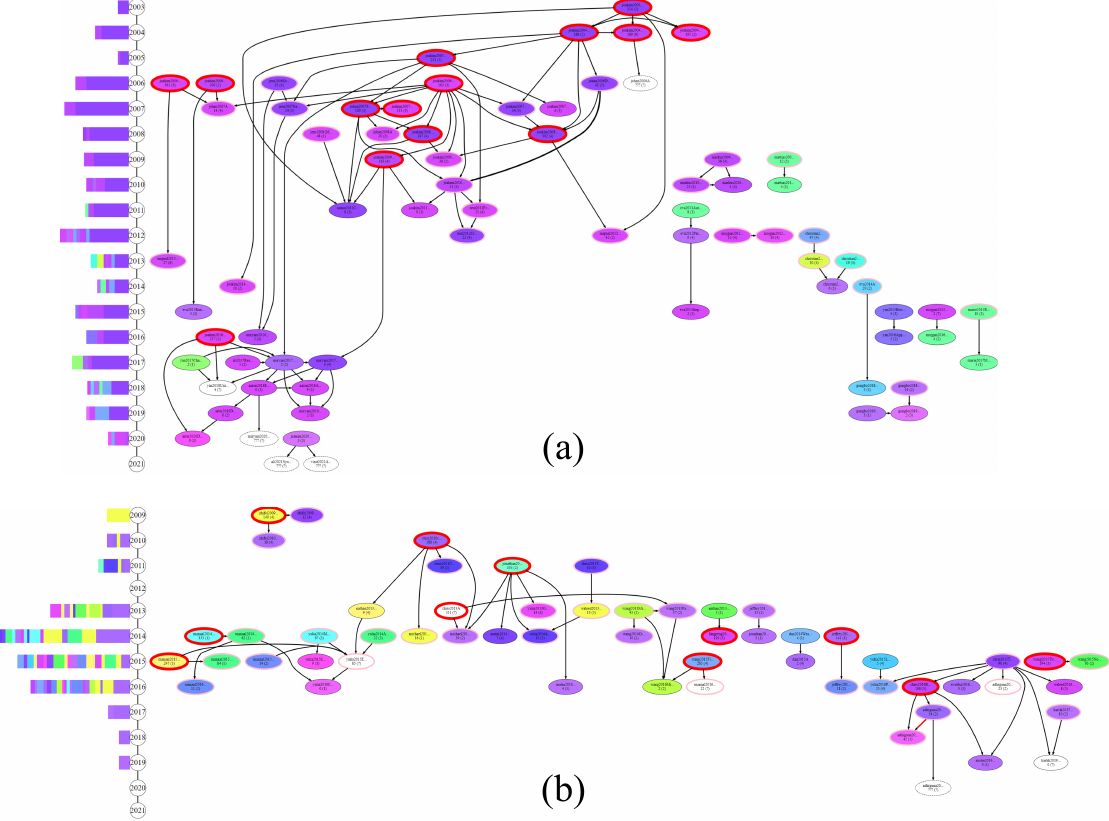}
\vspace{-0.33 in}
\caption{Core GF profile of representative scholars with similar h-index: (a) awardee; (b) non-awardee.}
\vspace{-0.20 in}
\label{fig:VisCaseHIndex}
\end{figure}


We have developed a visualization tool for GF profiling. Case studies are conducted with this tool to validate the effectiveness of GF for the understanding and analysis of scholar's impact. In a typical trial, we study two NLP scholars with similar h-index in the field. As shown in \rfig{VisCaseHIndex}(a), a representative NLP awardee presents a core GF profile that is moderate-sized and well-connected. Notably, s/he mostly worked on the same topic (in similar node fill color) during the entire career while having a local cluster of more than 10 high-citation papers (by red, thick node outline). In comparison, the other non-awardee generates a rather flat GF profile (\rfig{VisCaseHIndex}(b)). Most of his/her publications are within a short period of 4 years and the topics are quite diversified. More case study can be found in the video demo: \url{https://vimeo.com/795348791/}. A description of the visualization interface is also provided in Appendix~\ref{ap-c}.


In summary, the case study demonstrates that our GF profiling method can reveal key differences between award recipients and the other scholars having similar impact indicators. 
While the core profile of awardees are generally flawless with well-connected, sufficiently-sized graph, consistent topic development and moderate number of high-citation papers, the profile of non-awardees tend to suffer from certain ``deficiency'' for various reasons. 
Their profiles could be quite small, gap for many years, concentrate in short time period, contain very few high-citation paper, or involves too many research topics.
We caution that these distinctions are not definite and certain exceptions do exist. Indeed, award-receiving factors largely overlap with scientific impact but are not identical.


\vspace{-0.03 in}

\bsec{Conclusion}{Con}

This work presents GeneticFlow, a novel impact-oriented contextual scholar profiling method applied to big academic data. GF is equipped with three techniques to accomplish its design goal: an attributed graph representation using self-citation links to capture the scholar's scientific impact and evolution, a new unsupervised advisor-advisee detection algorithm for node profiling, and a well-engineered citation type classifier for edge profiling. Our method is evaluated on the real-world task of scientific award inference, for which a new benchmark dataset covering major CS sub-fields has been established. Experiment results demonstrate that the proposed GF method significantly outperforms existing indicator-based and network-based profiling methods. Further investigation on the variation of GF profiles reveals that a scholar's core papers and extend-type self-citations, i.e., the core GF profile, are the most important regarding his/her scientific impact. A rather compact core GF profile with 63.6\%$\sim$66.5\% nodes and 12.5\%$\sim$29.9\% edges suffices to undertake the award inference task in comparison to alternatives. Visualization of GF profiles also helps to identify key patterns of high-impact scholars in the studied research field.

\begin{acks}
This work was supported by National Key R\&D Program of China (2021YFB3500700), NSFC Grant 62172026, National Social Science Fund of China 22\&ZD153, and SKLSDE. 
\end{acks}
\clearpage

\bibliographystyle{ACM-Reference-Format}
\bibliography{GeneticFlow}


\appendix

\section{Alphabetical authorship detection}
\label{ap-aa}
We apply the hypothesis test to 8 CS sub-fields in our dataset. In each field, the papers authored by top-500 scholars are considered. 
As listed in \rtab{Authorship}, the test statistics computed theoretically, i.e., $p_{a-order}$ (null), vary between 0.197 and 0.356. The observed $p_{a-order}$ can be obtained by counting the number of papers following the alphabetical authorship in the dataset. 
By examining the bias of observed test statistics from those by null hypothesis (the last column in \rtab{Authorship}), it can be found that the fields of TCS and PL (bias = 0.147 $\sim$ 0.286) are quite different from the other fields (bias = 0.011 $\sim$ 0.118), with large increases in the observed $p_{a-order}$. These two fields are then excluded from GF profiling, also because of the large number of alphabetically ordered papers there. Note that a strict hypothesis test is not conducted here, due to the difficulty to estimate the variance and distribution of test statistics. However, a primitive bias analysis has effectively detected the CS sub-fields with significant use of alphabetical authorship (TCS and PL).

\begin{table}[t]
\centering
\setlength{\tabcolsep}{2pt}
\caption{Hypothesis test result for alphabetical authorship.}
\vspace{-0.13 in}
\label{tab:Authorship}
\begin{tabular}{@{}cccccc@{}}
\toprule
Field & \# of papers & \makecell{\# of papers \\ (a-order)} & \makecell{$p_{a-order}$ \\ (observation)}  &  \makecell{$p_{a-order}$ \\ (null)}  & bias\\ \midrule
NLP-ARC &26904 &7883 &0.293 &0.258 &0.035\\
Database &24559 &8847 &0.360 &0.242 &0.118\\
Security &35072 &11545 &0.329 &0.228 &0.101\\
DM &45050 &12030 &0.267 &0.197 &0.070\\
HCI &32731 &8992 &0.275 &0.252 &0.023\\
SE &23886 &8964 &0.375 &0.305 &0.070\\
TCS &33910 &19752 &0.582 &0.296 &0.286\\
PL &29966 &15088 &0.504 &0.356 &0.147\\
\bottomrule
\end{tabular}
\end{table}

\begin{figure}[t]
\centering
\includegraphics[width=0.85\linewidth]{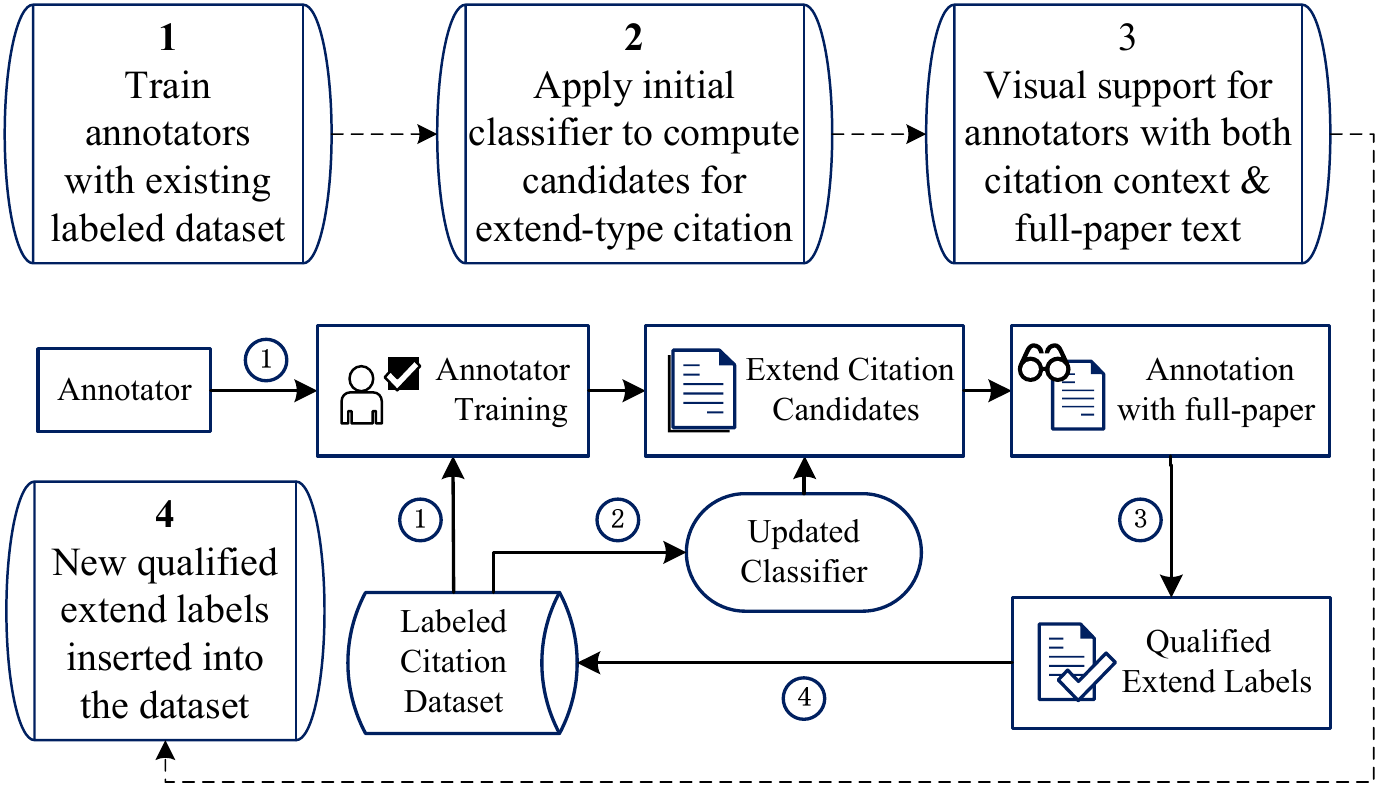}
\vspace{-0.18 in}
\caption{The annotation process for extend-type citations.}
\vspace{-0.15 in}
\label{fig:AnnotationTool}
\end{figure}

\begin{figure}[h]
\centering
\includegraphics[width=\linewidth]{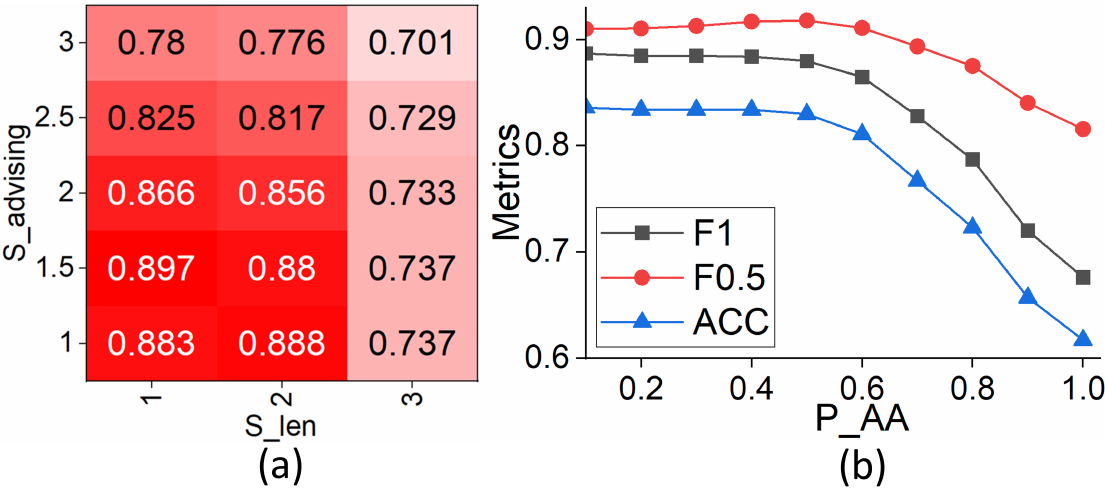}
\vspace{-0.35 in}
\caption{Advisor-advisee detection performance with varying parameters: (a) F1 over $S_{len}$, $S_{adr}$ ($p_{AA} \geq 0.5$ as the decision boundary); (b) F1/F0.5/ACC over detection boundaries of $p_{AA}$ ($S_{len} = 2$, $S_{adr} = 1.5$).}
\vspace{-0.1 in}
\label{fig:AAParameter}
\end{figure}

\begin{figure*}[t]
\centering
\includegraphics[width=\linewidth]{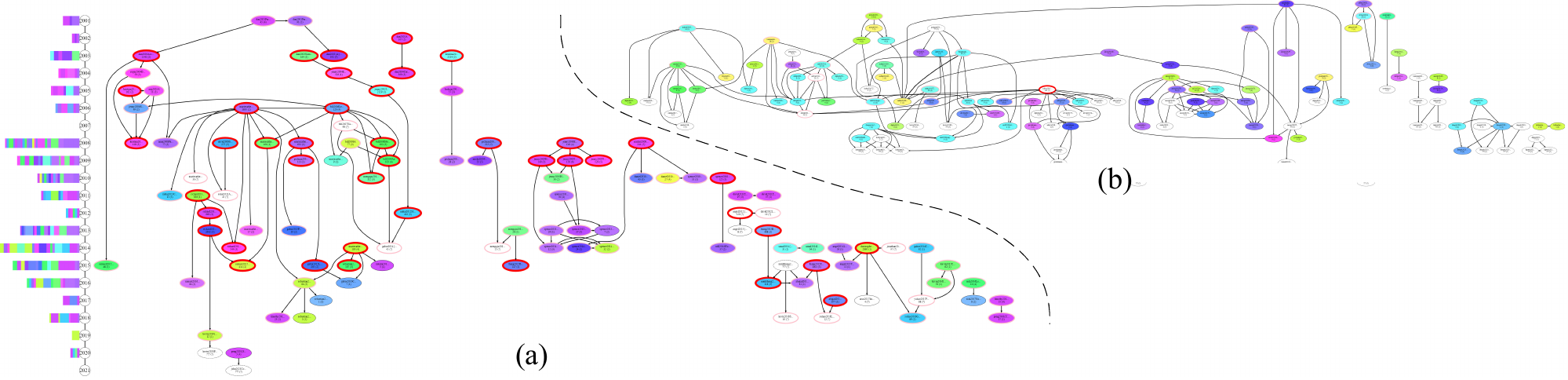}
\vspace{-0.3 in}
\caption{GF profiling examples of two NLP scholars with similar \# of papers: (a) awardee; (b) non-awardee.}
\vspace{-0.13 in}
\label{fig:VisCasePaperCount}
\end{figure*}

\section{Extend-Type Citation Annotation Process}
\label{ap-a}
The annotation job for extend-type citations is hard to proceed as they are rare in the data and difficult to identify even by human coders. To increase both the efficiency and accuracy of data annotations, we adopt an optimized process and tool support after insightful pilot study (\rfig{AnnotationTool}). First, each coder is required to go through all the 89 extend-citation labels from the previous data to understand the definition. Second, instead of starting from a random citation sample from NLP-ARC, the coders work on a set of high-potential candidate citations for extend types. The set is computed by applying the classifier trained on the initial labeled data with 89/1604 samples, from which 1000 candidate citations with relatively high extend-type probabilities are selected. Third, a visual interface is designed to ease the annotation process. For each candidate citation (citing-cited paper pair), coders will be provided a full list of the corresponding citation context. When the coder classifies it as the extend type, s/he can drill down to the full text of the citing paper for a double check. Lastly, the annotation process is partitioned into multiple batches. After each batch, the confirmed positive labels are added to the training data of the initial classifier, and the candidate citations are updated accordingly.

Three human coders with existing knowledge on citation classification research are recruited. For each positive label to be confirmed, all the three coders should agree on the extend-type citation. Finally, 133 new extend-type labels are obtained, leading to a dataset of 222/1604 samples.

\section{Advisor-Advisee Data Crawling from OpenReview}
\label{ap-b}
We extract real-life advisor-advisee datasets from OpenReview, a website hosting open peer review process. To access OpenReview service, scholars need to establish an academic profile page, which occasionally includes information about his/her advisor (name and period). As OpenReview mainly hosts peer reviews of AI venues, we consider the NLP and DM sub-fields in our work which largely overlap with AI research. From the NLP-ARC and DM datasets in \rtab{FieldEva}, we obtain top-200 scholars by h-index respectively. For each top scholar, all his/her papers in the sub-field are retrieved. On each paper, in case the top scholar is not the first author, one candidate advisor-advisee is assembled as a 3-tuple: $<$top scholar (advisor), first author (advisee), year of publication$>$. All candidate advisor-advisees are then verified on the OpenReview website to discover positive labels, negative labels, and unlabeled data.

To be labeled as a positive advisor-advisee pair, the first author should tag the top scholar on OpenReview as the advisor with a time period covering the year of publication. For negative advisor-advisees, the first author should never tag the top scholar as his/her advisor. In addition, the first author should tag at least another advisor in every year s/he co-authored with the top scholar. We adopt this strict rule for negative labels to guarantee accuracy: first, a large percentage of scholars do not tag any advisor on OpenReview, which leaves a possibility for positive labels; second, for any absent period of advisors, the top scholar could be the actual advisor, which potentially extends to the paper under investigation.

Finally, 365/116 and 145/35 positive/negative advisor-advisee labels are obtained from NLP-ARC and DM datasets respectively by the above strategy. The labeled data from NLP-ARC is then used to calibrate the parameters of the proposed advisor-advisee detection algorithm in \rsubsec{NodeRe}. As shown in \rfig{AAParameter}(a), we fix the advisor-advisee detection boundary as $p_{AA} \geq 0.5$ and draw the distribution of F1 metrics for detection with varying $S_{len}$ and $S_{adr}$. The distribution with other detection boundaries are similar. It can be observed that the F1 metrics outside the parameters of 2 will drop quickly with both $S_{len}$ and $S_{adr}$. Also consider that it is unrealistic to have $S_{len} = 1$ or $S_{adr} = 1$ (only one year or one co-authored paper during the advising). The final optimal parameters are set to $S_{len} = 2$ and $S_{adr} = 1.5$. Using these parameters, we further examine the choice of decision boundary for $p_{AA}$. \rfig{AAParameter}(b) shows that all the three metrics drop quickly after the threshold of 0.5. Also, we value precision of detection more than recall as the recall is already optimized by the definition. On the most important metric of F0.5, the threshold of 0.5 achieves the best score (0.918). Therefore, we use $p_{AA} \geq 0.5$ as the decision boundary for advisor-advisee detection.

\section{Description of Visualization Interface}
\label{ap-c}
We have developed a visualization interface for GF profiling. Compared with previous impact indicators using single, abstract metric, our graph-based profiling can help to visually discover detailed patterns on high-impact scholars. The visualization tool focuses on the NLP field where the NLP-ARC dataset is the most complete, so that the extracted GF profile are the most accurate. The GF profiles of top-ranked NLP scholars are visualized.

\end{document}